\documentclass[twocolumn,english]{revtex4}
\usepackage[T1]{fontenc}
\usepackage[latin9]{inputenc}
\usepackage{amsmath}
\usepackage{amssymb}
\usepackage{graphicx}

\makeatletter
\@ifundefined{textcolor}{}
{%
 \definecolor{BLACK}{gray}{0}
 \definecolor{WHITE}{gray}{1}
 \definecolor{RED}{rgb}{1,0,0}
 \definecolor{GREEN}{rgb}{0,1,0}
 \definecolor{BLUE}{rgb}{0,0,1}
 \definecolor{CYAN}{cmyk}{1,0,0,0}
 \definecolor{MAGENTA}{cmyk}{0,1,0,0}
 \definecolor{YELLOW}{cmyk}{0,0,1,0}
}

\usepackage{babel}

\makeatother

\usepackage{babel}
\begin{document}

\title{Geometrically frustrated Ising-Heisenberg spin model on expanded
Kagomé lattice}

\author{Onofre Rojas\thanks{correspondig email: ors@dfi.ufla.br; Phone: +5535 38291954.} }

\address{Departamento de Física, Universidade Federal de Lavras, CP 3037,
37200-000, MG, Brazil}
\begin{abstract}
Here we consider the Ising-Heisenberg model in the expanded Kagomé
lattice, also known as triangle-dodecagon (3-12) or star lattice.
This model can still be understood as a decorated honeycomb lattice.
Assuming that the Heisenberg spins are at the vertices of the triangle
while other spins are of the Ising type. Thus, this model is equivalent
to an effective Kagomé Ising lattice, through the decoration transformation
technique. Thus this means that the model is exactly solvable so we
can study the most relevant properties of this model. Like the phase
diagram at zero temperature, exhibiting a frustrated phase, a ferromagnetic
phase, a classical ferrimagnetic phase and a quantum ferrimagnetic
phase. We observed that Heisenberg spin exchange interaction influences
the frustrated phase, but we rigorously verify that the magnitude
and origin of the frustration emerge in a similar way to antiferromagnetic
Ising Kagomé lattice. Likewise, the thermodynamic properties of the
model can also be obtained, such as the critical temperature as a
dependence of the Hamiltonian parameters and the spontaneous magnetization
of the model. Besides, we investigated the entropy of the model, identifying
its residual entropy in the frustrated region. Even we analyze the
specific heat behavior as a temperature dependence, to deal with the
phase transition.
\end{abstract}
\maketitle

\section{Introduction}

One of the most relevant topics in statistical physics is the search
for exactly solvable models. In general, spin models in statistical
physics can not be solved exactly, so most of them can only be studied
numerically. Thus, the exact solutions were obtained only for limited
cases. After the solution found by Onsager for two-dimensional Ising
model\cite{onsager}, it inspired several attempts to solve other
similar models. Such as the honeycomb lattice\cite{Horiguchi-Wu,Kolesik},
whose exact solution of a honeycomb lattice with an external magnetic
field was provided by Wu\cite{Wu-jmp73}. Besides, Kagomé lattice was
also widely discussed in the literature \cite{azaria,lu-wu} and reference
there in. 

Typically geometric frustration arises in spin triangular structures.
When the competing antiferromagnetic interactions cannot be satisfied
simultaneously, leading to a considerable degeneracy of ground states.
In such a way, frustrated magnets have been attracted great scientific
interest because of quantum spin liquid in two-dimensional systems,
which has been proposed to play a striking role in high-temperature
superconductors. Furthermore, theoretical investigations have confirmed
that spin-1/2 Kagomé compounds are one of the natural candidates for
obtaining a quantum spin liquid ground state. This means, due to the
strong quantum fluctuation and geometric frustration would be responsible
for extinguishing the classical long-range magnetism at low temperature. 

The importance of investigations for geometrically frustrated kagomé
lattice compound is a great challenge. But there are few materials
with spin-1/2 kagomé structure that exhibit the quantum spin liquid
state in zero temperature. One of the typical illustrative examples
could be $\mathrm{Zn}\mathrm{Cu_{3}}\mathrm{(OH)_{6}}\mathrm{Cl_{2}}$\cite{shores,Han},
which shows a regular kagomé lattice. Using the single crystal sample
of $\mathrm{Zn}\mathrm{Cu_{3}}\mathrm{(OH)_{6}}\mathrm{Cl_{2}}$ were
observed several relevant characteristics of the quantum spin liquid
phase. Besides, investigations of spin-1/2 systems are quite relevant
in another context. Such discussed the magnetic properties through
canonical ensemble thermodynamic potentials of magnetic systems\cite{Plascak}.
 And the study of spin-1/2 bilayer system with the Glauber-type stochastic
dynamic behavior using the effective-field theory approach\cite{Ertas}.

Several decorated spin models can be transformed by applying the well-known
decoration transformation established in the 1950's by M. E. Fisher\cite{Fisher}
and Syozi\cite{syozi}. Later generalized in reference \cite{PhyscA-09,strecka-pla,Roj-sou11},
for arbitrary spins, such as the classical or quantum spin models.
This transformation is essential because we can map cumbersome models
into a simple or exactly solvable models. Below we mention few typical
examples where this approach was successfully applied. Geometrical
frustrated Cairo pentagonal lattice Ising model\cite{M-cairo}. The
Blume-Emery-Griffiths (BEG)\cite{BEG} model on the honeycomb lattice,
further investigated by Horiguchi\cite{Horiguchi-Wu}, Wu \cite{wu-pla86b},
Tucker\cite{tucker} and Urumov\cite{urumov-hnycmb}, applying the
standard decoration transformation\cite{Fisher,syozi} and satisfying
the Horiguchi's condition\cite{Horiguchi-Wu}. As well as XXZ-Ising
model on the triangular Kagomé lattice with spin-1/2, was studied
using analytical\cite{loh,strecka-triang} and Monte Carlo simulations\cite{loh}.

On the other hand low-dimensional square-hexagon (denoted for simplicity
by 4-6) Ising with spin-1/2 model was discussed by Lin and Chan\cite{Lin-4-6}
using the eight-vertex models mapping, later generalized as XXZ-Ising
model on a square-hexagon (4\textendash 6) lattice with spin-1/2,
which was investigated using the same approach\cite{our-4-6-latt}.
Similarly, the 3-12 lattice also known in the literature\cite{Lin-3-12,barry95},
as the star lattice, Fisher lattice, expanded Kagomé lattice, or even
triangular honeycomb lattice. Motivated the study of this kind of
model due to its closely relation with geometrically frustrated magnetic
material polymers\cite{octa-kagome}.

The present work is organized as follows. In Sec. 2, we present the
expanded Kagomé Ising-Heisenberg model. In Sec. 3, we discuss the
phase transition at zero temperature. While in Sec. 4, we give the
mapping of expanded Kagomé Ising-Heisenberg model into the Kagomé
Ising model. Furthermore, in Sec. 5, we discuss the thermodynamics
of the models, such as critical temperature, spontaneous magnetization
entropy, and specific heat. Finally, in sec. 6 we offer our conclusions.

\section{Expanded Kagomé lattice Hamiltonian}

\begin{figure}[h]
\includegraphics[scale=0.45]{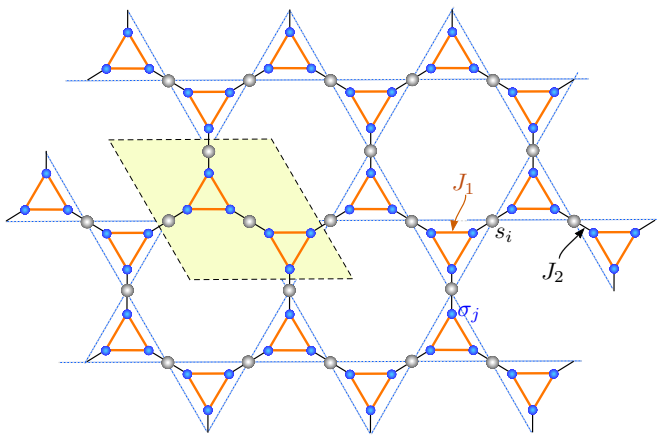}\caption{\label{fig:Two-exp-latt}Schematic lattice representation of expanded
Kagomé Ising-Heisenberg model. Small circles represent the Heisenberg
spins, while large circles denote the Ising spins, thick lines represent
the Heisenberg coupling while thin lines correspond to the Ising coupling.
The parallelogram is symbolizing the expanded Kagomé lattice unit
cell.}
\end{figure}

The Ising-Heisenberg expanded Kagomé lattice is built up by triangles
and dodecagons (3-12). Where thick solid line in triangles structure
between small circles represents the Heisenberg spin exchange bonds
(see Fig. \ref{fig:Two-exp-latt}). Likewise, the remaining couplings
(thin solid line) connecting small and large circles correspond to
Ising type exchange interactions as illustrated in Fig.\ref{fig:Two-exp-latt}.
This lattice can even be identified as a decorated star lattice\cite{octa-kagome,Zheng},
which synthesizes a chemical compound. Besides, Fig.\ref{fig:Two-exp-latt}
illustrates the unit cell of expanded Kagomé lattice by a parallelogram.
 Thus, each unit cell is composed of 3 Heisenberg spins and 6 Ising
spins. 

The Hamiltonian that describes the expanded Kagomé lattice can be
expressed by 
\begin{equation}
H=-J_{1}\sum_{<i,j>}(\boldsymbol{\sigma}_{i},\boldsymbol{\sigma}_{j})_{\Delta}-J_{2}\sum_{<k,l>}\sigma_{k}^{z}s_{l},\label{eq:Ham-1}
\end{equation}
here by $\sigma^{\alpha}$ we denote the Heisenberg spins operator
with $\alpha=\{x,y,z\}$, whereas $s_{i}$ denotes the Ising spin
$s_{i}=\pm1/2$. The first summation corresponds to the anisotropic
Heisenberg coupling between the nearest neighbor, which is explicitly
given by
\begin{equation}
J_{1}(\boldsymbol{\sigma}_{i},\boldsymbol{\sigma}_{j})_{\Delta}=J_{1}\left(\sigma_{i}^{x}\sigma_{j}^{x}+\sigma_{i}^{y}\sigma_{j}^{y}\right)+\Delta\sigma_{i}^{z}\sigma_{j}^{z},
\end{equation}
 where $J_{1}$ is the Heisenberg exchange interaction in $xy$ component
and $\Delta$ correspond to Heisenberg exchange interaction in $z$
component.

While the second summation represents the sum of the nearest Ising
and Heisenberg spins coupling and $J_{2}$ denotes the Ising spin
exchange interaction parameter. 

\section{Phase diagram}

\begin{figure}
\includegraphics[scale=0.35]{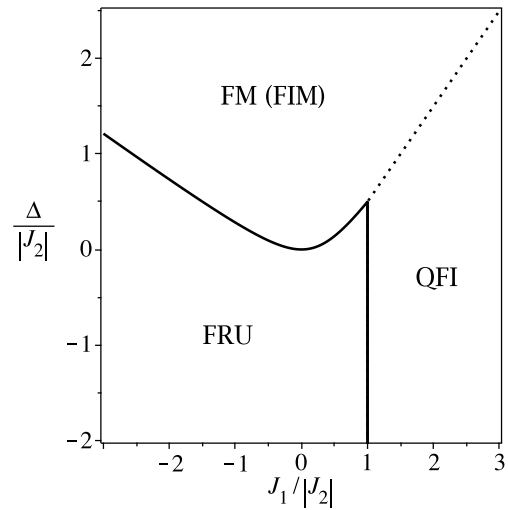}\caption{\label{fig:Phase-diagram}Zero temperature phase diagram in the plane
$J_{1}/|J_{2}|$ - $\Delta/|J_{2}|$.}
\end{figure}

The phase diagram at zero temperature is illustrated in units of $|J_{2}|$
as shown in Fig.\ref{fig:Phase-diagram}. Thus the diagram exhibits
a ferromagnetic (FM) phase for $J_{2}>0$ and a ferrimagnetic (FIM)
phase for $J_{2}<0$, with the corresponding ground state energy per
unit cell given by 
\begin{equation}
E=-\frac{3|J_{2}|}{2}-\frac{3\Delta}{2},\text{ with }\begin{cases}
FM; & \;\text{for}\;J_{2}>0\\
FIM; & \;\text{for}\;J_{2}<0
\end{cases}.
\end{equation}

\begin{figure}
\includegraphics[scale=0.44]{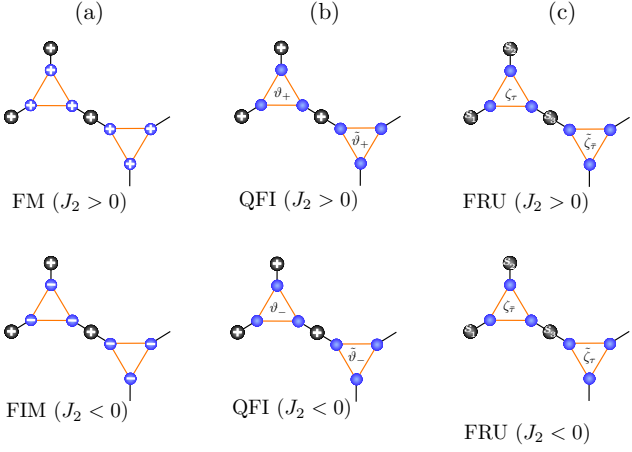}\caption{\label{fig:configuration}Ground state configurations per unit cell.
(a) Classical Ferromagnetic (Ferrimagnetic), (b) quantum ferrimagnetic
(QFI) phase and in (c) frustrated (FRU) phase. }
\end{figure}

This ground state configuration per unit cell is schematically depicted
in Fig.\ref{fig:configuration}a, where the corresponding eigenstates
can also be expressed as follows
\begin{alignat}{1}
|FM\rangle= & |_{+}^{+}\rangle\otimes|_{+}^{+}\hspace{-0.12cm}\vartriangleright\hspace{-0.15cm}\begin{smallmatrix}+\end{smallmatrix}\rangle\otimes|+\rangle\otimes|\begin{smallmatrix}+\end{smallmatrix}\hspace{-0.16cm}\vartriangleleft_{+}^{+}\rangle,\label{eq:FM}\\
|FIM\rangle= & |_{+}^{+}\rangle\otimes|_{-}^{-}\hspace{-0.12cm}\vartriangleright\hspace{-0.15cm}\begin{smallmatrix}-\end{smallmatrix}\rangle\otimes|+\rangle\otimes|\begin{smallmatrix}-\end{smallmatrix}\hspace{-0.16cm}\vartriangleleft_{-}^{-}\rangle.\label{eq:FIM}
\end{alignat}
Clearly, \eqref{eq:FM} and \eqref{eq:FIM} are invariant under total
spin inversion. 

The total magnetization per unit cell of expanded Kagomé lattice is
given by
\begin{equation}
m_{t}=\frac{m_{I}+2m_{H}}{3},\label{eq:tot-mg}
\end{equation}
with $m_{I}$ being the Ising spins magnetization per Ising spin sites,
and $m_{H}$ denotes the Heisenberg spins magnetization per Heisenberg
spin sites.

Using this setting, the Ising spin magnetization is given by $m_{I}=1/2$
and the Heisenberg spin magnetization is $m_{H}=1/2$. Therefore,
for $J_{2}>0$ arises the FM phase (here we denote $s=\sigma=+$)
with a total magnetization per unit cell is $m_{t}=1/2$.

Whereas for $J_{2}<0$ the system is in FIM phase ($s=-\sigma=+$,
by $+$ we mean $+1/2$) with Ising spin magnetization $m_{I}=1/2$,
while Heisenberg spin magnetization is $m_{H}=-1/2$ and the total
magnetization per unit cell becomes $m_{t}=-1/6$.

The other state illustrated in Fig.\ref{fig:configuration} is the
quantum ferrimagnetic (QFI) phase, which is represented schematically
in Fig.\ref{fig:configuration}b. This state can be written explicitly
as follows,
\begin{equation}
|QFI\rangle=|_{s}^{s}\rangle\otimes|\vartheta_{\sigma}\rangle\otimes|s\rangle\otimes|\tilde{\vartheta}_{\sigma}\rangle,\label{eq:QFI}
\end{equation}
with 
\begin{alignat}{1}
|\vartheta_{\sigma}\rangle= & \tfrac{1}{\sqrt{3}}\left(|_{\sigma}^{\sigma}\hspace{-0.12cm}\vartriangleright\hspace{-0.15cm}\begin{smallmatrix}\bar{\sigma}\end{smallmatrix}\rangle+|_{\sigma}^{\bar{\sigma}}\hspace{-0.12cm}\vartriangleright\hspace{-0.15cm}\begin{smallmatrix}\sigma\end{smallmatrix}\rangle+|_{\bar{\sigma}}^{\sigma}\hspace{-0.12cm}\vartriangleright\hspace{-0.15cm}\begin{smallmatrix}\sigma\end{smallmatrix}\rangle\right),\\
|\tilde{\vartheta}_{\sigma}\rangle= & \tfrac{1}{\sqrt{3}}\left(|\begin{smallmatrix}\bar{\sigma}\end{smallmatrix}\hspace{-0.16cm}\vartriangleleft_{\sigma}^{\sigma}\rangle+|\begin{smallmatrix}\sigma\end{smallmatrix}\hspace{-0.16cm}\vartriangleleft_{\bar{\sigma}}^{\sigma}\rangle+|\begin{smallmatrix}\sigma\end{smallmatrix}\hspace{-0.16cm}\vartriangleleft_{\sigma}^{\bar{\sigma}}\rangle\right),
\end{alignat}
here, by $\bar{\sigma}$ we mean $\bar{\sigma}=-\sigma$. A particular
case of QFI state becomes
\begin{equation}
|QFI\rangle=\begin{cases}
|_{+}^{+}\rangle\otimes|\vartheta_{+}\rangle\otimes|+\rangle\otimes|\tilde{\vartheta}_{+}\rangle, & J_{2}>0\\
\\
|_{+}^{+}\rangle\otimes|\vartheta_{-}\rangle\otimes|+\rangle\otimes|\tilde{\vartheta}_{-}\rangle, & J_{2}<0
\end{cases}.
\end{equation}
Since the corresponding ground state energy is given by 
\begin{align}
E_{QFI}= & -\frac{|J_{2}|}{2}+\frac{\Delta}{2}-2J_{1}.
\end{align}
 In QFI phase, there are also two possible total magnetizations: (i)
When $J_{2}>0$ the spin configuration satisfy $s=\sigma=+$, whose
Ising spin magnetization $m_{I}=1/2$ and Heisenberg spin magnetization
$m_{H}=1/6$ and its corresponding total magnetization is $m_{t}=5/18$.
(ii) When $J_{2}<0$ the spin configuration satisfy $s=-\sigma$,
thus the Ising spin magnetization $m_{I}=1/2$, Heisenberg spin magnetization
$m_{H}=-1/6$ and the total magnetization for this configuration is
given by $m_{t}=1/18$.

In Fig.\ref{fig:Phase-diagram} is also reported a frustrated (FRU)
phase, at zero temperature. A schematic spin configuration displayed
in Fig. \ref{fig:configuration}c, whose state is expressed below
\begin{equation}
|FRU\rangle=\begin{cases}
|_{s_{2}}^{s_{1}}\rangle\!\otimes\!|\zeta_{\tau}\rangle\!\otimes\!|s_{3}\rangle\!\otimes\!|\tilde{\zeta}_{\bar{\tau}}\rangle, & J_{2}>0\\
\\
|_{s_{2}}^{s_{1}}\rangle\!\otimes\!|\zeta_{\bar{\tau}}\rangle\!\otimes\!|s_{3}\rangle\!\otimes\!|\tilde{\zeta}_{\tau}\rangle, & J_{2}<0
\end{cases},\label{eq:fru-st}
\end{equation}
and denoting the Heisenberg spins states by
\begin{alignat}{1}
|\zeta_{\tau}\rangle= & \tfrac{1}{\sqrt{2+a^{2}}}\left(a\,|_{\sigma_{2}}^{\sigma_{1}}\hspace{-0.12cm}\vartriangleright\hspace{-0.15cm}\begin{smallmatrix}\sigma_{3}\end{smallmatrix}\rangle+|_{\sigma_{1}}^{\sigma_{3}}\hspace{-0.12cm}\vartriangleright\hspace{-0.15cm}\begin{smallmatrix}\sigma_{2}\end{smallmatrix}\rangle+|_{\sigma_{3}}^{\sigma_{2}}\hspace{-0.12cm}\vartriangleright\hspace{-0.15cm}\begin{smallmatrix}\sigma_{1}\end{smallmatrix}\rangle\right),\\
|\tilde{\zeta}_{\tau}\rangle= & \tfrac{1}{\sqrt{2+a^{2}}}\left(a\,|\begin{smallmatrix}\sigma_{1}\end{smallmatrix}\hspace{-0.16cm}\vartriangleleft_{\sigma_{2}}^{\sigma_{3}}\rangle+|\begin{smallmatrix}\sigma_{3}\end{smallmatrix}\hspace{-0.16cm}\vartriangleleft_{\sigma_{1}}^{\sigma_{2}}\rangle+|\begin{smallmatrix}\sigma_{2}\end{smallmatrix}\hspace{-0.16cm}\vartriangleleft_{\sigma_{3}}^{\sigma_{1}}\rangle\right).
\end{alignat}
Where Ising spins are restricted to $s_{1}+s_{2}+s_{3}=\tau$, analogously
we have $\sigma_{1}+\sigma_{2}+\sigma_{3}=\tau$, with $\tau$ restricted
to $\tau=\pm$ (here $\pm1/2$), and defining for convenience $\bar{\tau}=-\tau$.
The frustrated state \eqref{eq:fru-st} illustrated in Fig. \ref{fig:configuration}c
are equivalent through Heisenberg spin inversion.

Whereas the coefficients is given by 
\begin{equation}
a=\begin{cases}
u_{_{1,+}}=-\frac{1}{2}+\frac{J_{2}}{J_{1}}+\frac{J_{-}}{2J_{1}}, & J_{2}>0\\
\\
u_{_{2,+}}=-\frac{1}{2}-\frac{J_{2}}{J_{1}}+\frac{J_{+}}{2J_{1}}, & J_{2}<0
\end{cases}.
\end{equation}
with $J_{_{\pm}}=\sqrt{4J_{2}^{2}+9J_{1}^{2}\pm4J_{1}J_{2}}$ (denoted
just for convenience).

The corresponding ground state energy of a frustrated state is given
by

\begin{align}
E_{FRU}=\begin{cases}
-\frac{J_{2}}{2}+\frac{\Delta-J_{1}-J_{_{-}}}{2}, & J_{2}>0\\
\\
\frac{J_{2}}{2}+\frac{\Delta-J_{1}-J_{_{+}}}{2}, & J_{2}<0.
\end{cases} & .
\end{align}

It is worth to mention that the magnetizations of Ising spins and
Heisenberg spins in the frustrated region are null. So the only one
responsible for generating frustration is Ising spins, as discussed
in Ref. \cite{kano53}.

A straight line $\Delta/|J_{2}|=-\frac{1}{2}+J_{1}/|J_{2}|$ gives
the boundary between FM(FIM) phase and QFI phase. It deserves to remark
that the interface between (FIM or FM) and QFI only occur at zero
temperature. While the boundary between FM(FIM) phase and FRU phase
is described by the curve $\frac{\Delta}{|J_{2}|}=-\frac{1}{2}+\frac{J_{1}+J_{_{-}}}{4|J_{2}|}$,
and this phase transition persist at finite temperature which will
be discussed later. Afterward, the phase transition between QFI and
FRU is plainly given by a $J_{1}/|J_{2}|=1$, this interface also
remains at finite temperature.

\section{Expanded Kagomé lattice on Ising-Heisenberg model}

\begin{figure}[h]
\begin{centering}
\includegraphics[scale=0.8]{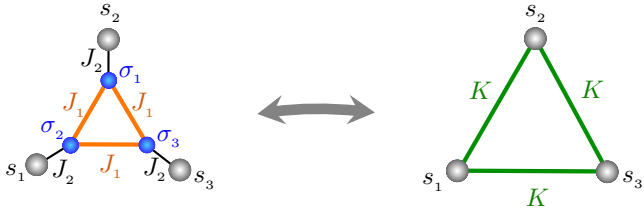} 
\par\end{centering}
\caption{\label{fig:3-leg-hybrid-star}Decoration transformation of 3-leg star
hybrid-spin and triangle Ising spin.}
\end{figure}

Now let us consider a three-leg hybrid-star system as schematically
depicted  in Fig.\ref{fig:3-leg-hybrid-star}(left), where the three
Heisenberg spins localized at the vertex of a triangle corresponding
to the system decoration. In which the internal bond $J_{1}$ is of
the Heisenberg-type interaction since the outer legs are of Ising-type
interaction $J_{2}$ as shown in Fig. \ref{fig:3-leg-hybrid-star}(left
side). This system can be mapped in a triangle with the Ising spin
coupling $K$ through a direct decoration transformation approach\cite{Roj-sou11}
or using the standard decoration transformation\cite{Fisher,syozi,strecka-pla,PhyscA-09},
which is shown in Fig. \ref{fig:3-leg-hybrid-star}(right). 

Let us define the following operator 
\begin{equation}
\mathbf{V}(\{\boldsymbol{\sigma},s\})=\exp\left\{ \beta\sum_{i=1}^{3}\left[J_{1}(\boldsymbol{\sigma}_{i},\boldsymbol{\sigma}_{i+1})_{\Delta}+J_{2}\sigma_{i}^{z}s_{i}\right]\right\} .
\end{equation}

Thus the Boltzmann factors of decorated hybrid-spin model become 
\begin{equation}
w(\{s\})=\mathrm{tr}_{\{\sigma\}}\Bigl(\mathbf{V}(\{\boldsymbol{\sigma},s\})\Bigr),
\end{equation}
the index of spin $s_{l}$ in Hamiltonian \eqref{eq:Ham-1}, was re-indexed
by $s_{i}$ (same index of $\sigma_{i}$ ) just for simplicity imposing
no restriction.

The inner triangle system (decorated) is expressed as Heisenberg coupling,
and it provides two configurations for Ising spins (legs), these correspond
to the following configurations $\{\uparrow\uparrow\uparrow\}$ and
$\{\uparrow\uparrow\downarrow\}$. So defining $\varsigma=s_{1}+s_{2}+s_{3}$,
the pair configurations become $\varsigma=3/2$ and $1/2$. Therefore,
we get the following Boltzmann factors defined by $w(1/2)=w_{1}$
and $w(3/2)=w_{3}$, and then we find that 
\begin{align}
w_{1}= & \left(z^{3}\!+\!z^{-1}x^{-2}\right)\!\left(y\!+\!y^{-1}\!\right)\!+\!xz^{-1}\!\left(y^{-1}v_{_{+}}\!+y\,v_{_{-}}\!\right)\!,\label{eq:w(1/2)}\\
w_{3}= & z^{3}\left(y^{3}+y^{-3}\right)+z^{-1}\left(x^{4}+2x^{-2}\right)\left(y+y^{-1}\right).\label{eq:w(3/2)}
\end{align}
Here we have introduced the following notations $x=\mathrm{e}^{\frac{\beta J_{1}}{4}}$,
$y=\mathrm{e}^{\frac{\beta J_{2}}{4}}$, $z=\mathrm{e}^{\frac{\beta\Delta}{4}}$
and $v_{_{\pm}}=2\cosh\left(\tfrac{\beta J_{_{\pm}}}{4}\right)$,
just as a shorthand way of writing the Boltzmann factors.

Now let us assume that the Hamiltonian of effective Kagomé lattice
Ising model can be expressed as follows
\begin{equation}
\mathcal{H}_{eff}=-K_{0}-K\sum_{<i,j>}s_{i}s_{j},\label{eq:H-eff}
\end{equation}
where $K_{0}$ is ``constant'' energy and $K$ is effective coupling
parameter of the Kagomé lattice Ising model, while the summation runs
over nearest neighbor interactions.

After carrying out the direct decoration transformation\cite{Roj-sou11},
the effective Kagomé lattice Ising model also has only the same couple
of configurations $\{\uparrow\uparrow\uparrow\}$ and $\{\uparrow\uparrow\downarrow\}$,
which correspond to $\varsigma=3/2$ and $\varsigma=1/2$, respectively.
Therefore, both models must be equivalent. That means $\tilde{w}(\varsigma)=w(\varsigma)$,
we have two algebraic equations with two unknown parameters $K_{0}$
and $K$, thus we are able to solve the algebraic system equations,
\begin{align}
\mathrm{e}^{\beta K_{0}}\exp\left(-\beta K/4\right)= & w_{1}=\tilde{w}_{1},\\
\mathrm{e}^{\beta K_{0}}\exp\left(3\beta K/4\right)= & w_{3}=\tilde{w}_{3}.
\end{align}

Thereafter, the unknown parameters in the effective Kagomé Ising model
could be expressed in terms of all arbitrary parameters of the expanded
Kagomé Ising-Heisenberg spin model {[}see Eq. \eqref{eq:Ham-1}{]},
\begin{align}
K= & \frac{1}{\beta}\ln\left(\tfrac{w_{3}}{w_{1}}\right),\label{eq:3-legs-K}\\
K_{0}= & \frac{1}{4\beta}\ln\left(w_{1}^{3}w_{3}\right),\label{eq:3-legs-K0}
\end{align}
where $w_{1}$ and $w_{3}$ are given by \eqref{eq:w(1/2)} and \eqref{eq:w(3/2)}
respectively.

\section{Thermodynamics}

Now we are going to study the thermodynamics of the present model,
so we need to get the free energy per unit cell of the model that
can be written as
\begin{equation}
f_{EK}=-2K_{0}+f_{K}.
\end{equation}
With $f_{K}$ being the effective Kagomé lattice Ising model free
energy\cite{syozi,kano53,azaria-prl87} and $K_{0}$ corresponds to
the effective \textquotedbl{}constant\textquotedbl{} energy of the
effective Kagomé lattice. The factor 2 in $K_{0}$ comes from unit
cell, note that we have two decorated systems per each unit cell.

The free energy of effective Kagomé lattice per unit cell \cite{syozi,kano53,azaria-prl87}
can be expressed using a single integral\cite{fan-wu-70}, as follows
\begin{equation}
f_{K}=-\frac{T}{4\pi}\int_{0}^{2\pi}\ln\left[A(\phi)+\sqrt{Q(\phi)}\right]\mathrm{d}\phi,
\end{equation}
where $A(\phi)$ and $Q(\phi)$ are defined by
\begin{alignat*}{1}
A(\phi)= & \tfrac{1}{2}r^{4}+9r^{2}+12r+\tfrac{21}{2}-2(r-1)\left(r^{2}-1\right)\cos(\phi),\\
Q(\phi)= & A(\phi)^{2}-8(r-1)^{2}(r^{2}-1)^{2}\left[1+\cos(\phi)\right],
\end{alignat*}
with $r=\frac{w_{3}}{w_{1}}.$ Thus, the free energy per unit cell
of expanded Kagomé lattice becomes
\begin{equation}
f_{EK}\!=\!-2T\ln\negthickspace\left(w_{1}\right)\negmedspace-\negmedspace\frac{T}{4\pi}\negthickspace\int_{0}^{2\pi}\negthickspace\ln\negthickspace\left[\negmedspace A(\phi)\negmedspace+\negmedspace\sqrt{Q(\phi)}\right]\negthickspace\mathrm{d}\phi.\label{eq:fek}
\end{equation}

Before continuing studying the thermodynamics properties, we need
to remark three interesting points. 

First concerning to a residual entropy in FRU region, occurs when
$T\rightarrow0$ and $w_{3}<w_{1}$, this implies that $r=\frac{w_{3}}{w_{1}}\rightarrow0$,
then the elements of integral reduce to 
\begin{alignat}{1}
A(\phi)= & 2\cosh(\phi)+\tfrac{21}{2},\label{eq:Ap}\\
Q(\phi)= & \left[2\cosh(\phi)+\tfrac{21}{2}\right]^{2}+8\left[1+\cos(\phi)\right],\label{eq:Qp}
\end{alignat}
 both functions are independent of $T$ and $r$. Therefore, the free
energy \eqref{eq:fek} can be solved numerically for \eqref{eq:Ap}
and \eqref{eq:Qp}, getting $f_{EK}\approx-1.50549949\,T$; thus,
the residual entropy becomes $\mathcal{S}\approx1.50549949$, as expected
independent of temperature. This result complies with the frustrated
region found in reference \cite{kano53}. We recognized that Heisenberg
spin exchange interaction influences the frustrated (FRU) phase, but
the origin of frustration and the magnitude arise in a similar way
to antiferromagnetic Ising Kagomé lattice\cite{kano53}.

Second, the condition for $\frac{w_{3}}{w_{1}}=r=1$ and $T\rightarrow0$,
occurs in the interface of QFI and FRU states, so we have $A(\phi)=2^{5}$
and $Q(\phi)=2^{10}$; thus the free energy reduces to 
\begin{equation}
f_{EK}=-2T\ln\left(w_{1}\right)-\frac{T}{4\pi}\int_{0}^{2\pi}\ln\left(2^{6}\right)\mathrm{d}\phi,\label{eq:fek-1}
\end{equation}
this implies that the residual entropy merely becomes as $\mathcal{S}=3\ln(2)\approx2.07944$.

Third, when $w_{3}>w_{1}$ and $T\rightarrow0$, then $r\rightarrow\infty$,
so we have
\begin{alignat}{1}
A(\phi)\approx & 2r^{3}\cos(\phi)+\tfrac{1}{2}r^{4}\sim\tfrac{1}{2}r^{4},\\
Q(\phi)\approx & \tfrac{1}{4}r^{8}-8r^{6}(1+\cos(\phi))\sim\tfrac{1}{2}r^{8}.
\end{alignat}
Consequently, the free energy reduces to
\begin{alignat}{1}
f_{EK}\!\approx & \!-2T\!\ln\!\left(w_{1}\right)\negmedspace-\negmedspace\frac{T}{4\pi}\negthickspace\int_{0}^{2\pi}\negthickspace\ln\negthickspace\left(\negmedspace\tfrac{1}{2}r^{4}\negmedspace+\negmedspace\sqrt{\tfrac{1}{4}r^{8}}\right)\negthickspace\mathrm{d}\phi,\nonumber \\
\approx & -2T\ln\negthickspace\left(w_{1}\right)\negmedspace-2T\negmedspace\ln\negthickspace\left(\frac{w_{3}}{w_{1}}\right),\nonumber \\
\approx & -2T\ln\negthickspace\left(w_{3}\right).
\end{alignat}
In addition, as expected, there is no frustration in this region FM(FIM)
or QFI because there is no residual entropy ($\mathcal{S}=0$ ) when
$T=0$.

Once the free energy is obtained, we can analyze the thermodynamic
properties of the expanded Kagomé Ising-Heisenberg model.

\subsection{Critical temperature}

In the following, we will discuss one of the essential properties
of the two-dimensional lattice models, the critical behavior of temperature.
It is well established that the critical temperature for the Kagomé
lattice is given by $\frac{K}{T_{c}}=\ln\left(3+2\sqrt{3}\right)$
\cite{syozi,barry}. As an alternative, we are able to write the critical
temperature as a ratio of Boltzmann factor:
\begin{equation}
\frac{w_{3}^{c}}{w_{1}^{c}}=r_{c}=3+2\sqrt{3},\label{eq:crit-T}
\end{equation}
where $w_{1}^{c}$ and $w_{3}^{c}$ refer to the Boltzmann factors
at a critical temperature, which is given by \eqref{eq:w(1/2)} and
\eqref{eq:w(3/2)} at $T=T_{c}$, respectively.

After some algebraic manipulation, this condition can be rewritten
more explicitly as
\begin{equation}
z_{c}^{4}=\frac{\left(x_{c}^{6}+2\right)(3+2\sqrt{3})-1-\frac{y_{c}^{-1}v_{_{+}}^{c}+y_{c}v_{_{-}}^{c}}{x_{c}^{-3}\left(y_{c}+y_{c}^{-1}\right)}}{x_{c}^{2}\left(1-\left(y_{c}^{2}-1+y_{c}^{-2}\right)(3+2\sqrt{3})\right)}.
\end{equation}

\begin{figure}
\includegraphics[scale=0.85]{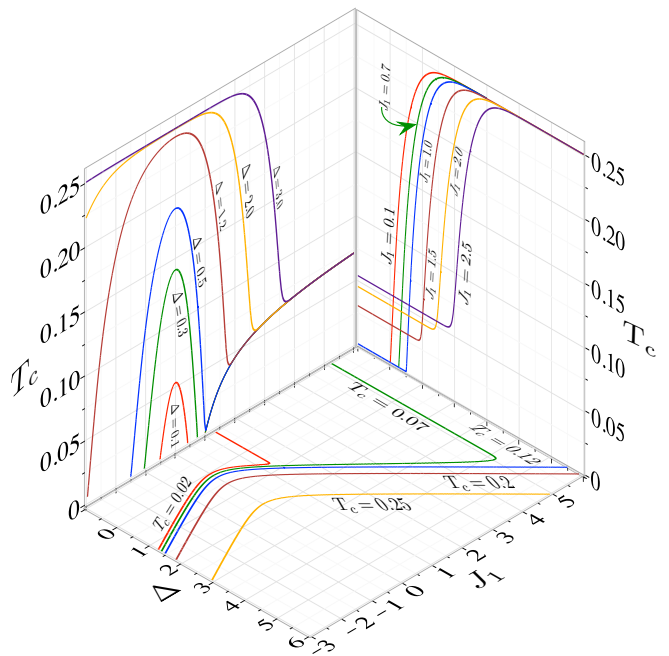}\caption{\label{fig:crit-temp}(Left panel) Critical temperature $T_{c}$ against
$J_{1}$, for several values of $\Delta$. (Right panel) Critical
temperature $T_{c}$ as a function of $\Delta$, for a range of values
in $J_{1}$. (Bottom) The phase diagram in the plane $\Delta-J_{1}$
for several critical temperature values. In all panels a fixed $J_{2}=1$
was considered.}
\end{figure}

In Fig.\ref{fig:crit-temp}(left) is illustrated the critical temperature
$T_{c}$ as a function of $J_{1}$ for several values of $\Delta$
and fixing $J_{2}=1$. The dashed line corresponds to $\Delta=0.5$,
which is a special curve with two critical points at zero temperature
(for detail see Fig.\ref{fig:Phase-diagram}). For larger value of
$\Delta>0.5$ only occurs one critical point at zero temperature,
while for $0<\Delta<0.5$, there are three critical points at zero
temperature. However, for $\Delta<0$ only appears one critical point
at zero temperature, which always occurs at $J_{1}=1$ and is independent
of $\Delta$. This curve agrees with the zero temperature phase diagram
shown in Fig. \ref{fig:Phase-diagram}.

Similarly, Fig.\ref{fig:crit-temp}(right) shows the critical temperature
$T_{c}$ as a function of $\Delta$ and for several values of $J_{1}$.
For $\Delta<0$, the critical temperature is independent of $\Delta$.
Likewise, for $\Delta\rightarrow\infty$, the critical temperature
also becomes independent of $\Delta$, and in this limit, we have
$T_{c}\rightarrow0.25$. While, in Fig.\ref{fig:crit-temp}(bottom)
the phase diagram at fixed critical temperature $T_{c}$ in the $J_{1}-\Delta$
plane is reported and for different critical temperature $T_{c}$
values. When $T_{c}\rightarrow0$, this phase diagram leads to a zero
temperature phase diagram (see the solid line in Fig.\ref{fig:Phase-diagram}),
since there is no evidence of phase transition between FM and QFI
phase at zero temperature. It is worth noting that, for $J_{2}=-1$,
the curves become identical, the only difference is that, instead
of FM phase we have an FIM phase.

Using this expression, we can obtain the critical temperature in the
limit of $J_{1}/|J_{2}|\rightarrow\infty$, leading to
\begin{equation}
T_{c}=\tfrac{|J_{2}|}{6\ln\left(2+\sqrt{3}+\sqrt{6+4\,\sqrt{3}}\right)}=0.0836826082|J_{2}|.
\end{equation}

While for $\Delta/|J_{2}|\rightarrow\infty$, the critical temperature
leads to $T_{c}/|J_{2}|=0.25$.

\subsection{Spontaneous Magnetization }

Another relevant quantity to analyze here is the spontaneous magnetization
of the present model. Therefore, it is pertinent to examine the magnetization
of the expanded Kagomé Ising-Heisenberg lattice. The Ising spin magnetization
$m_{I}=\langle s\rangle$, can be determined concerning the effective
Kagomé Ising lattice magnetization, while the magnetization of Heisenberg
spins can be obtained using the decoration transformation approach\cite{Fisher,syozi,PhyscA-09,Roj-sou11}
\begin{equation}
\langle\sigma_{1}^{z}\rangle=\eta\langle s_{1}\rangle+\frac{\gamma}{3}\langle s_{1}s_{2}s_{3}\rangle,
\end{equation}
 which linearly combines single and triple Ising spin average $\langle s\rangle$
and $\langle s_{1}s_{2}s_{3}\rangle$.

To find the coefficients $\eta$ and $\gamma$ we use the following
relation\cite{Fisher,syozi,PhyscA-09,Roj-sou11}
\begin{equation}
\zeta(\{s\})=\bigl[\eta(s_{1}+s_{2}+s_{3})+\gamma s_{1}s_{2}s_{3}\bigr]w(\{s\}).\label{eq:Zeta-o}
\end{equation}
In a similar way, we define
\begin{equation}
\tilde{\zeta}(\{s\})=\mathrm{tr}_{\{\sigma\}}\left[\left(\sum_{i=1}^{3}\sigma_{i}^{z}\right)\mathbf{V}(\{\boldsymbol{\sigma},s\})\right].
\end{equation}

On the other hand, the spin coupling configurations $\{\uparrow\uparrow\uparrow\}$
and $\{\uparrow\uparrow\downarrow\}$, is denoted merely as $\tilde{\zeta}(1/2)=\tilde{\zeta}_{1}$
and $\zeta(3/2)=\zeta_{3}$. Therefore, we have
\begin{alignat}{1}
\tilde{\zeta}_{3}= & \mathrm{tr}_{\{\sigma\}}\left[\mathbf{U}_{3}\left(\sum_{i=1}^{3}\sigma_{i}^{z}\right)\mathbf{U}_{3}^{-1}\mathbf{D}_{3}\right],\label{eq:zeta3}\\
\tilde{\zeta}_{1}= & \mathrm{tr}_{\{\sigma\}}\left[\mathbf{U}_{1}\left(\sum_{i=1}^{3}\sigma_{i}^{z}\right)\mathbf{U}_{1}^{-1}\mathbf{D}_{1}\right],\label{eq:zeta1}
\end{alignat}
where $\mathbf{D}_{3}$ and $\mathbf{D}_{1}$ are the diagonalized
matrix representation of $\mathbf{V}$ for each sector, and these
are given by
\begin{alignat}{1}
\mathbf{D}_{3}={\rm diag}\{ & \tfrac{3}{2}y^{3}z^{3},\tfrac{-1}{2zyx^{2}},\tfrac{-1}{2zyx^{2}},\tfrac{x^{4}}{2zy},\tfrac{-x^{4}}{2zy},\nonumber \\
 & \tfrac{y}{2zx^{2}},-\tfrac{3}{2}y^{-3}z^{3}\}
\end{alignat}
and

\begin{alignat}{1}
\mathbf{D}_{1}={\rm diag}\{ & \tfrac{3}{2}yz^{3},\tfrac{1}{2zyx^{2}},\tfrac{xy}{2zv_{_{-}}},\tfrac{xyv_{_{-}}}{2z},-\tfrac{x}{2zyv_{_{+}}},\tfrac{-xv_{_{+}}}{2zy},\nonumber \\
 & \tfrac{y}{2zx^{2}},-\tfrac{3}{2}y^{-1}z^{3}\}.
\end{alignat}
The orthogonal matrices $\mathbf{U}_{3}$ and $\mathbf{U}_{1}$ are
obtained straightforwardly for each configuration $\{\uparrow\uparrow\uparrow\}$
and $\{\uparrow\uparrow\downarrow\}$ respectively, which are expressed
below

\begin{equation}
\mathbf{U}_{3}=\left(\begin{array}{cccccccc}
1 & 0 & 0 & 0 & 0 & 0 & 0 & 0\\
0 & 0 & 0 & 1 & 0 & -1 & -\frac{1}{2} & 0\\
0 & 0 & 0 & 1 & 0 & 1 & -\frac{1}{2} & 0\\
0 & -1 & -\frac{1}{2} & 0 & 1 & 0 & 0 & 0\\
0 & 0 & 0 & 1 & 0 & 0 & 1 & 0\\
0 & 0 & 1 & 0 & 1 & 0 & 0 & 0\\
0 & 1 & -\frac{1}{2} & 0 & 1 & 0 & 0 & 0\\
0 & 0 & 0 & 0 & 0 & 0 & 0 & 1
\end{array}\right),
\end{equation}
and

\begin{equation}
\mathbf{U}_{1}=\left(\begin{array}{cccccccc}
1 & 0 & 0 & 0 & 0 & 0 & 0 & 0\\
0 & 0 & u_{_{1,-}} & u_{_{1,+}} & 0 & 0 & 0 & 0\\
0 & -1 & 1 & 1 & 0 & 0 & 0 & 0\\
0 & 0 & 0 & 0 & -1 & 1 & 1 & 0\\
0 & 1 & 1 & 1 & 0 & 0 & 0 & 0\\
0 & 0 & 0 & 0 & 1 & 1 & 1 & 0\\
0 & 0 & 0 & 0 & 0 & u_{_{2,-}} & u_{_{2,+}} & 0\\
0 & 0 & 0 & 0 & 0 & 0 & 0 & 1
\end{array}\right),
\end{equation}
where $u_{1,\pm}=-\frac{1}{2}+\frac{J_{2}}{J_{1}}\pm\frac{J_{-}}{2J_{1}}$
and $u_{2,\pm}=-\frac{1}{2}-\frac{J_{2}}{J_{1}}\pm\frac{J_{+}}{2J_{1}}$.

Consequently, the relations \eqref{eq:zeta3} and \eqref{eq:zeta1}
after some algebraic manipulation, simply becomes as 
\begin{alignat}{1}
\tilde{\zeta}_{3}= & \tfrac{3}{2}z^{3}\left(y^{3}-y^{-3}\right)+z^{-1}\left(\tfrac{1}{2}x^{4}+x^{-2}\right)\left(y-y^{-1}\right),\label{eq:tZ3}\\
\tilde{\zeta}_{1}= & \left(y-y^{-1}\right)\left(\tfrac{3}{2}z^{3}-\tfrac{1}{2}x^{-2}z^{-1}\right)\nonumber \\
 & +\tfrac{1}{2}\left(v_{_{-}}y-v_{_{+}}y^{-1}\right)xz^{-1}.\label{eq:tZ1}
\end{alignat}

From Eq. \eqref{eq:Zeta-o} we also have the following relations

\begin{alignat}{1}
\zeta_{3}= & \left(\frac{3}{2}\eta+\frac{1}{8}\gamma\right)w_{3},\\
\zeta_{1}= & \left(\frac{1}{2}\eta-\frac{1}{8}\gamma\right)w_{1}.
\end{alignat}
From where we obtain the unknown coefficients
\begin{alignat}{1}
\eta= & \frac{1}{2}\left(\frac{\zeta_{3}}{w_{3}}+\frac{\zeta_{1}}{w_{1}}\right),\\
\gamma= & 2\left(\frac{\zeta_{3}}{w_{3}}-3\frac{\zeta_{1}}{w_{1}}\right),
\end{alignat}
where $\zeta_{3}=\tilde{\zeta}_{3}$ and $\zeta_{1}=\tilde{\zeta}_{1}$,
with $\tilde{\zeta}_{3}$ and $\tilde{\zeta}_{1}$ given by Eqs. \eqref{eq:tZ3}
and \eqref{eq:tZ1}.

Using the relations for the correlation functions obtained by Barry
et al.\cite{barry} (see Eq.(2.9a) of Ref. \cite{barry}). Now we
need to relate the three spin thermal average and single spin average
\begin{equation}
\langle s_{1}s_{2}s_{3}\rangle=R(r)\,\langle s_{1}\rangle,
\end{equation}
where $R(r)$ after some algebraic manipulation becomes 
\begin{equation}
R(r)=\frac{r^{3}-3r^{2}-r-5}{4(r-1)^{3}}.
\end{equation}
Moreover, the single Heisenberg spins average can be written as
\begin{equation}
\langle\sigma^{z}\rangle=\left[\eta+\frac{\gamma}{3}R(r)\right]\langle s_{1}\rangle.
\end{equation}
On the other hand, the thermal average of Ising spin $\langle s_{1}\rangle$\cite{barry}
is given by 
\begin{equation}
\langle s_{1}\rangle=\left(r^{2}-6\,r-3\right)^{\frac{1}{8}}G(r),
\end{equation}
where $G(r)$ is defined as in reference \cite{barry}, For the present
model $G(r)$ has been adapted as a function of $r$, which is expressed
as

\begin{equation}
G(r)=\left(\tfrac{r+3}{r-1}\right)^{\frac{1}{4}}\tfrac{\left(r^{2}+2\,r+5\right)^{\frac{3}{8}}}{2\left(r+1\right)}.
\end{equation}
We emphasize that $G(r)$ is always a positive amount. Furthermore,
the factor $2$ in the denominator is included because we are considering
the Ising spin eigenvalues as $\pm1/2$. For our case, the thermal
average of Ising spin and Heisenberg spin, are defined as $m_{I}=\langle s\rangle$
and $m_{H}=\langle\sigma\rangle$ respectively.

Notice that the magnetization exponent satisfies the same universality
class of that Kagomé lattice (same critical exponent 1/8).

Hence, the total magnetization per spin can be expressed using the
Eq.\eqref{eq:tot-mg}, so we have

\begin{equation}
m_{t}=\frac{\langle s\rangle+2\langle\sigma^{z}\rangle}{3}.
\end{equation}

\begin{figure}
\includegraphics[scale=0.45]{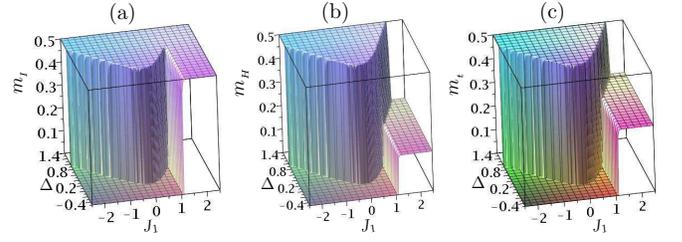}\caption{\label{fig:Magn3d}Magnetization as a function of $J_{1}$ and $\Delta$,
assuming $T=0.01$ and $J_{2}=1$. (a) Ising spin magnetization. (b)
Heisenberg spin magnetization. (c) Total spin magnetization per unit
lattice.}

\end{figure}

In Fig.\ref{fig:Magn3d} is illustrated the magnetization of Ising
spin as a function of $J_{1}$ and $\Delta$, considering fixed parameter
$J_{2}=1$ and temperature $T=0.01$. Fig.\ref{fig:Magn3d}a displays
the Ising spin magnetization, and we observe the regions FM and QFI
have the magnetization leading to $\langle s\rangle\rightarrow0.5$,
whereas in FRU region the magnetization becomes null. In the interface
between FM and QFI, there is no spontaneous magnetization for Ising
spin, this means there is no phase transition at finite temperature,
but there is only a zero temperature phase transition.

Analogously Fig.\ref{fig:Magn3d}b displays Heisenberg spin magnetization,
here we can see clearly the magnetization in region FM leads to $\langle s\rangle\rightarrow0.5$,
while the magnetization in QFI region becomes $\langle s\rangle\rightarrow1/6$,
and naturally, in FRU region the magnetization becomes null. Whereas
the total magnetization is depicted in Fig.\ref{fig:Magn3d}c: thus
in region FM $m_{t}\rightarrow0.5$, in QFI region $m_{t}\rightarrow1/3$
and for FRU region the total magnetization readily is null.

\subsection{Entropy and specific heat}

To complete our investigation concerning this model, let us discuss
the thermodynamic properties, which can be obtained straightforwardly
from the free energy $f_{EK}$, such as entropy $\mathcal{S}=-\partial f_{EK}/\partial T$,
internal energy $U=f_{EK}+T\mathcal{S}$ and specific heat $C=T\partial\mathcal{S}/\partial T$. 

\begin{figure}
\includegraphics[scale=0.5]{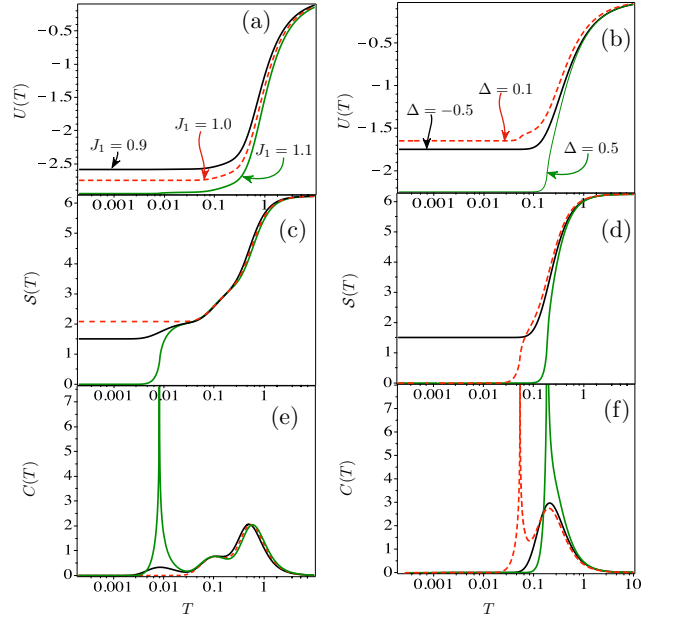}\caption{\label{fig:7}In left panel is considered fixed parameters $\Delta=-0.5$
and $J_{1}$ is given in (a). In right panel is assumed fixed $J_{1}=0.0$
and $\Delta$ is given in (b). (a-b) Internal energy as a function
of temperature for fixed parameters. (c-d) Entropy as a function of
temperature. (e-f) Specific heat as a dependence of temperature. }
\end{figure}

In Fig.\ref{fig:7}a is displayed the internal energy as a function
of temperature considering fixed parameter $J_{1}$ as described in
the panel (a): for $J_{1}=0.9$ and $J_{1}=1.0$ the curves increases
smoothly, but $J_{1}=1.1$ there is a tiny jump at $T_{c}\approx0.082$.
In Fig.\ref{fig:7}b is depicted the internal energy as a dependence
of temperature, for $\Delta=0.5$ and $\Delta=0.1$ there is a strong
change of curvature at $T_{c}\approx0.186$ and $T_{c}\approx0.0537$
respectively, whereas for $\Delta=-0.5$ there is no evidence of phase
transition at finite temperature. The next panel (c) in Fig.\ref{fig:7}
reports the entropy for the same set of parameters as in (a), for
$J_{1}=0.9$ the residual entropy leads to $\mathcal{S}=1.5055$,
while for $J_{1}=1.0$ (equivalently $r=1$) also exhibits another
residual entropy given by $\mathcal{S}=3\ln(2)=2.07944$, but for
$J_{1}=1.1$ there is no residual entropy because the system leads
to QFI phase. Similarly, in Fig.\ref{fig:7}d we show the entropy
as a function of temperature for the set of parameters of the panel
(b). Once again for $\Delta=0.5$ and $\Delta=0.1$, there is a sudden
change close to the critical temperature, but for $\Delta=-0.5$ the
entropy increases smoothly indicating the absence of phase transition.
Finally, in Fig.\ref{fig:7}f confirms the critical temperature for
the same set of parameters described in panel (b) as a divergence
in specific heat.

\section{Conclusions}

Several materials have exotic structures such as honeycomb lattice,
a triangular lattice, Kagomé lattice among others. Hence it is worth
to investigate the magnetic properties and geometric frustration of
these kinds of models. Therefore, in this work was considered the
Ising-Heisenberg model on expanded Kagomé lattice, that we could also
name as triangle-dodecagon (3-12) lattice, this model can even be
viewed as a decorated star lattice. Considering all spins located
in triangles as Heisenberg spins while remaining spins are of Ising
type spins. Consequently, this model is equivalent to an effective
Kagomé Ising lattice, through a direct decoration transformation technique\cite{Roj-sou11}.
This model allows us to study zero temperature magnetic properties
of the 3-12 lattice, such as the phase diagram at zero temperature,
where we found four phases, a frustrated (FRU) phase, a ferromagnetic
(FM) phase, a classical ferrimagnetic (FIM) phase and a quantum ferrimagnetic
(QFI) phase. We remarked that Heisenberg spin exchange interaction
strongly influences the frustrated phase; however, we rigorously verified
that the magnitude and origin of the frustration turn out in the same
fashion as that of the antiferromagnetic Ising Kagomé lattice.\cite{kano53}.
We also obtain the free energy of the model which permit us to explore,
the critical temperature and the spontaneous magnetization were also
considered as a dependence of Hamiltonian parameters. In addition,
we have investigated the entropy where we observed a residual entropy
in the frustrated region $\mathcal{S}\approx1.5055$. As well as in
the interface between QFI and FRU the residual entropy is given by
$\mathcal{S}=3\ln(2)$. Besides, we also studied the specific heat
divergence as a function of temperature to handle the phase transition.

\section*{Acknowledgments}

This work was partially supported by Brazilian agencies FAPEMIG, CAPES
and CNPq.


\begin{thebibliography}{10}
\bibitem{onsager}L. Onsager, Phys. Rev. \textbf{65}, 117 (1944).

\bibitem{Horiguchi-Wu}T. Horiguchi, Phys. Lett. A \textbf{113}, 425
(1986) .

\bibitem{Kolesik} M. Kolesik and L. Samaj, Int. J. Mod. Phys. B \textbf{6},
1529 (1992).

\bibitem{Wu-jmp73}F. Y. Wu, J. Math. Phys. \textbf{15}, 687 (1974);
J. Phys. A: Math Gen. \textbf{23}, 375 (1990).

\bibitem{azaria}P. Azaria and H. Giacomini, J. Phys. A: Math. Gen.
\textbf{21,} L935 (1988).

\bibitem{lu-wu}W. T. Lu and F. Y. Wu, Phys. Rev. E \textbf{71}, 046120
(2005).

\bibitem{shores}Shores, M. P.; Nytko, E. A.; Bartlett, B. M.; Nocera,
D. G. J. Am. Chem. Soc. \textbf{127}, 13462 (2005)

\bibitem{Han}Han, T. H.; Helton, J. S.; Chu, S. Y.; Nocera, D. G.;
Rodriguez- Rivera, J. A.; Broholm, C.; Lee, Y. S. Nature \textbf{492},
406 (2012)

\bibitem{Plascak}J. A. Plascak, V.K. Varma, Journ. Magn. and Mag.
Mat. \textbf{468}, 224 (2018)

\bibitem{Ertas}M. Erta\c{s}, E. Kantar, M. Keskin, Journ. Magn. and
Mag. Mat. \textbf{358}, 56 (2014)

\bibitem{Fisher}M. E. Fisher, Phys. Rev. \textbf{113}, 969 (1959).

\bibitem{syozi}I. Syozi, in \textit{Phase Transitions and Critical
Phenomena}, edited by C. Domb and M. S. Green (Academic Press, New
York, 1972), Vol. 1.

\bibitem{PhyscA-09}O. Rojas, J. S. Valverde and S. M. de Souza, Physica
A \textbf{388}, 1419 (2009).

\bibitem{strecka-pla}J. Strecka, Phys. Lett. A, \textbf{374}, 3718
(2010) ; LAP LAMBERT Academic Publishing, Saarbrucken, Germany, 2010,
ISBN: 978-3-8383-6200-7. 

\bibitem{Roj-sou11}O. Rojas and S. M. de Souza, J. Phys. A: Math
Theor. \textbf{44}, 245001 (2011).

\bibitem{M-cairo}M. Rojas, O. Rojas, S. M. de Souza, Phys. Rev. E
\textbf{86}, 051116 (2012).

\bibitem{BEG}M. Blume, V. J. Emery and R. B. Griffiths, Phy. Rev.
A \textbf{4,} 1071 (1971).

\bibitem{wu-pla86b}F.Y. Wu, Phys. Lett. A \textbf{116}, 245 (1986).

\bibitem{tucker}J. W. Tucker, J. Appl. Phys. \textbf{69}, 6164 (1991).

\bibitem{urumov-hnycmb}V. Urumov, J. Phys. C: Solid State Phys. \textbf{20},
L875 (1987).

\bibitem{loh}Y. L. Loh, D. X. Yao and E. W. Carlson, Phys. Rev. B
\textbf{77}, 134402 (2008).

\bibitem{strecka-triang}J. Strecka, L. Canova, M. Jascur and M. Hagiwara,
Phys. Rev. B \textbf{78}, 024427 (2008); J. Cisarova, J. Strecka
Phys. Rev. B, \textbf{87} (2013), 024421; J. Cisarova, F. Michaud, F. Mila, J. Strecka, J. Phys. Rev. B. \textbf{87} (2013) 054419.

\bibitem{Lin-4-6}K. Y. Lin and T. L. Chen, Chin. J. Phys. \textbf{25},
178 (1987).

\bibitem{Lin-3-12}K. Y. Lin and J. L. Chen, J. Phys. A: Math. Gen.
\textbf{20}, 5695 (1987).

\bibitem{barry95}J. H. Barry and M. Khatun Phys. Rev. B \textbf{51},
5840 (1995).

\bibitem{our-4-6-latt}J. S. Valverde, O. Rojas, and S. M. de Souza,
Phys. Rev. E \textbf{79}, 041101 (2009).

\bibitem{octa-kagome}Y. Tang, C. Peng, W. Guo, J. Wang, G. Su, and
Z. He, J. Am. Chem. Soc. \textbf{139}, 14057 (2017).

\bibitem{Zheng}Y. Z. Zheng, M. L. Tong, W. Xue, W. X. Zhang, X.-M.
Chen, F. Grandjean, and G. J. Long, Angew Chem Int Ed Engl \textbf{46},
6076 (2007).

\bibitem{kano53}K. Kano and S. Naya, Prog. Theor. Phys. \textbf{10},
158 (1953).

\bibitem{azaria-prl87}P. Azaria, H. T. Diep and H. Giacomini, Phys.
Rev. Lett., \textbf{59},1629 (1987) .

\bibitem{fan-wu-70}C. Fan and F. Wu, Phys. Rev. B \textbf{2}, 723
(1970).

\bibitem{barry}J. H. Barry, T. Tanaka, M. Khatun and C. H. Munera,
Phys. Rev. B \textbf{44} 2595 (1991). 
\end{thebibliography}
\end{document}